\documentclass[twocolumn, showkeys, showpacs, amsmath, amssymb]{revtex4}

\usepackage[dvips]{graphicx}

\newcommand{\be}{\begin{equation}}

\newcommand{\diff}{{\mathrm d}}
\newcommand{\ee}{\end{equation}}
\newcommand{\eg}{{\it e.g.}, }
\newcommand{\Eq}[1]{Eq.~(\ref{#1})}
\newcommand{\eq}[1]{eq.~(\ref{#1})}
\newcommand{\fig}[1]{fig.~\ref{#1}}
\newcommand{\Fig}[1]{Fig.~\ref{#1}}
\newcommand{\Int}{I_n^\mathrm{t}}
\newcommand{\iave}{\langle i \rangle}
\newcommand{\ie}{{\it i.e.}, }
\newcommand{\ityp}{i^\mathrm{t}}
\newcommand{\LN}{\mathrm{LN}}

\newcommand{\oper}[2]{\mathrel{\mathop{\kern 0pt#1}\limits_{#2}}}
\newcommand{\Section}[1]{Section~\ref{#1}}

\newcommand{\vs}{{\it vs} }

\begin{document}

\title{Statistical properties of currents flowing through tunnel junctions}

\author{V. Da Costa}%
\affiliation{Institut de Physique et Chimie des Mat\'eriaux de
Strasbourg, CNRS (UMR 7504) and Universit\'e Louis Pasteur, 23 rue
du Loess, 67037 Strasbourg Cedex, France.}
\altaffiliation{Corresponding author: victor@ipcms.u-strasbg.fr, tel. 
33 (0)3 88 10 70 84, fax 33 (0)3 88 10 72 49}

\author{M. Romeo}%
\affiliation{Institut de Physique et Chimie des Mat\'eriaux de
Strasbourg, CNRS (UMR 7504) and Universit\'e Louis Pasteur, 23 rue
du Loess, 67037 Strasbourg Cedex, France.}

\author{F. Bardou}%
\affiliation{Institut de Physique et Chimie des Mat\'eriaux de
Strasbourg, CNRS (UMR 7504) and Universit\'e Louis Pasteur, 23 rue
du Loess, 67037 Strasbourg Cedex, France.}

\begin{abstract}
This paper presents an overview of the statistical properties
arising from the broadness of the distribution of tunnel currents
in metal-insulator-metal junctions. Experimental current
inhomogeneities can be modelled by a lognormal distribution and
the size dependence of the tunnel current is modified at small
sizes by the effect of broad distributions.
\end{abstract}

\pacs{73.40.Rw, 73.40.Gk, 05.40.Fb, 75.70.Pa, 05.60.-k}
\keywords{Tunnel junctions,  broad distributions, 
metal-insulator-metal structures, giant magnetoresistance, 
lognormal distribution}

\maketitle


\section{Introduction: MIM junctions and broad distributions}
\label{s1}

    Metal-Insulator-Metal (MIM) tunnel junctions have been introduced
into the physics toolbox four decades ago \cite{FiG61}. They have
given rise to several landmarks in condensed matter physics such
as the Josephson effect and the Coulomb blockade. Since the
discovery of large room temperature Tunnel Magneto-Resistance
(TMR) \cite{MiT95b,MKW95}, MIM junctions have been under intense
scrutiny again. This paper will summarize recent studies of the
disorder effects in MIM junctions, which is one of the aspects of
MIM junctions' physics. This topic is not new \cite{Cho63,Hur66}
but it can be revisited thanks to recent experimental and
theoretical developments.

Practioners know how difficult it is to achieve {\it
reproducibility} of the conductances of MIM junctions, even when
the junctions are prepared on the same wafer. This is becoming a 
crucial problem with the prospect of applications of TMR to 
Magnetic Random Access Memories and magnetic read heads, which 
require the conductance dispersion to be typically less than 10\%. 
This raises the question of whether the conductance 
irreproducibility is a purely technical problem or
whether there is something more fundamental behind it.

The observed large dispersion of conductances from one junction to
another one is statistically unusual and this provides a clue on
the nature of the problem. Consider, for instance, a $10\times
10$~$\mu$m$^2$ junction with a typical interatomic distance of
0.3~nm, so that the cross-section contains $n\simeq 10^{9}$~atoms.
According to the central limit theorem, relative fluctuations of
an ensemble of $n$ components scale as $1/\sqrt{n}$. Thus,
fluctuations of 10\% at the junction scale would correspond to
fluctuations of 3000 at the atomic scale. This suggests
that either the distribution of tunnel currents is extremely
broad, or that fluctuations do not average out as in
the central limit theorem, or both.

During the last fifteen years, the importance of such broad
distributions has emerged in several areas of statistical physics
related mostly to anomalous diffusion \cite{BoG90,SZF95,BBA02}.
The paradigm of broad distributions is the L\'evy flight, \ie
random walks in which the length $l$ of the free flight has a
power law distribution $P(l) = \alpha l_0^\alpha / l^{1+\alpha}$
($l>l_0$) with a diverging second moment ($0<\alpha <2$). With
such distributions, the variance is infinite, the usual central
limit theorem does not apply and the relative fluctuations of a
sum of $n$ terms do {\it not} decrease with the number of terms.
This reminds of the large fluctuations observed even in large
junctions. Moreover, the sum of $n$ terms displacements tends to
be dominated by a few of them which reminds of the infamous 'hot
spots', \ie of filamentary like structures carrying most of the
current. At last, with L\'evy flights, even the law of large
numbers can fail to apply, \ie the typical sum of $n$ terms might
not be proportional to $n$ even at large $n$ (case $\alpha\leq
1$). This suggests that the tunnel current might not always be
proportional to the size of the junction.

There seems to be a connection between broad distributions and MIM
tunnel junctions. To clarify the matter, one needs to know
experimentally the distribution of tunnel conductances
(\Section{s3}). Then one can study the consequences of the current
distribution (\Section{s4}), in particular the scale effects
(\Section{s5}).

\section{Experimental distribution of tunnel currents}
\label{s3}

Conducting Atomic Force Microscopy (C-AFM) can map the tunnel
current flowing through an oxide barrier \cite{DBB98}. In this
technique, the conducting tip of an atomic force microscope is
scanned in contact with the surface of the barrier, while a bias
voltage between the tip and the bottom metallic electrode creates
a current flowing through the barrier. In this way, one records
simultaneously the topography and the tunnel current. The actual
resolution of the C-AFM is difficult to estimate. Correlations
studies show that current structures smaller than 1 nm$^2$ are
resolved. There must, however, be some sort of convolution by the
finite tip size, which produces some {\it underestimation} of the
current inhomogeneities.

\begin{figure}
\includegraphics[scale=0.33]{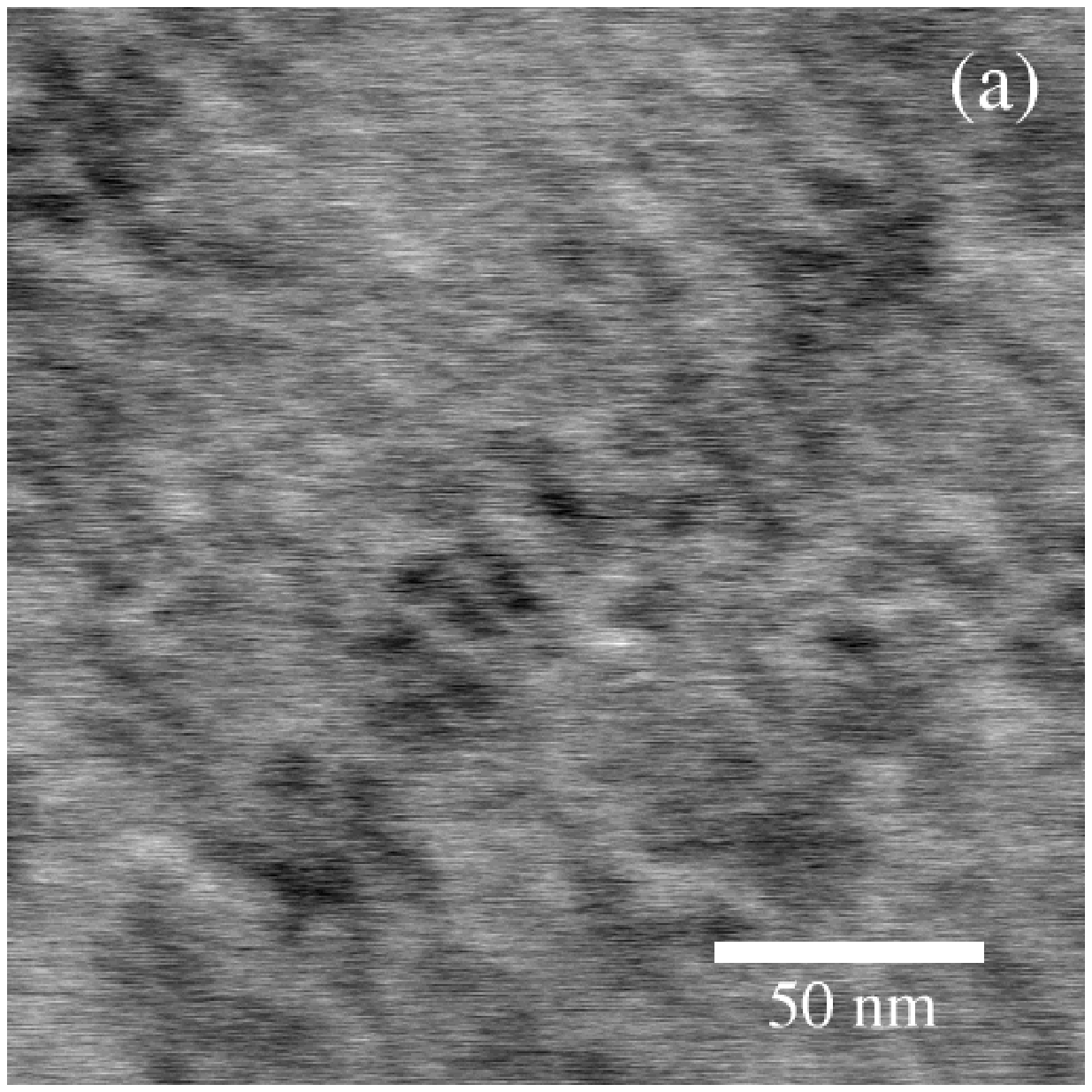}
\includegraphics[scale=0.33]{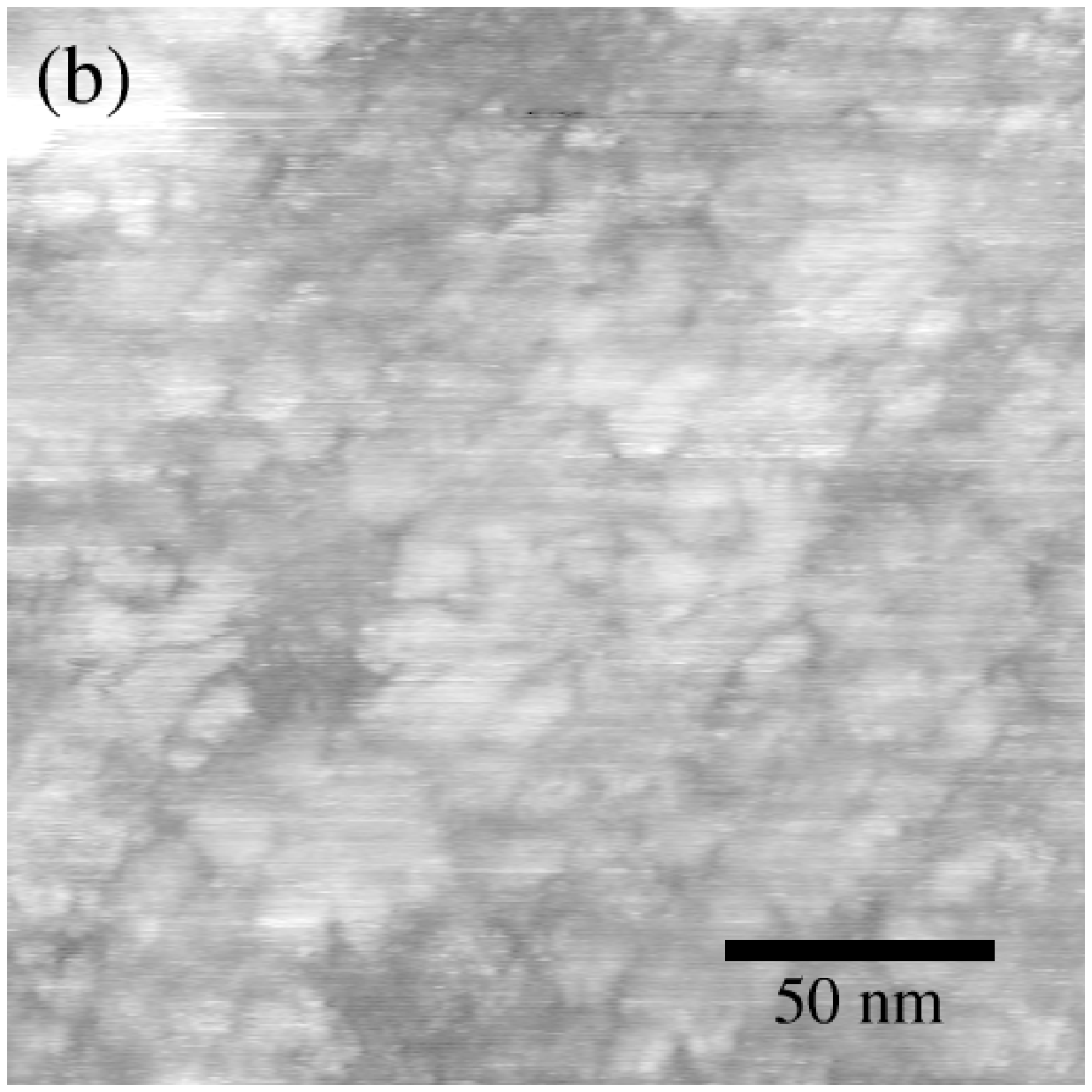}
\caption{(a) Topography (black = 0 nm, white = 0.2 nm) and (b) 
tunnel current (log scale, black = 40 pA, white = 1 nA)
images for an Al oxide barrier (Al deposited on Co, 
Ar + O$_2$ plasma oxidized, AlO$_x$ thickness $\simeq$ 1 nm, see \cite{DTD00}).}
\label{fig1}
\end{figure}

The topography and current images for a typical oxide (AlO$_x$)
barrier are shown in \fig{fig1}. The topography (\ref{fig1}a) is very smooth
(roughness $\simeq 0.2$~nm), whereas
the logarithm of the current image (\ref{fig1}b) exhibits a continuum of
current values going from $\simeq$~40~pA to $\simeq$~1~nA. 
The
local $I-V$ characteristics obtained with C-AFM are consistent
with tunnelling with transmission much less
than 1, even at the highest current points which are, thus, 
not pinholes. This type of
experiments has been reproduced by several groups with
similar results \cite{AKK99,AKK00,WKU01,AHK01,OBB01,LWP01}. They
are especially useful for barrier optimization, for instance with
respect to oxidation \cite{DTD00,DDT01} or annealing
\cite{AKH00,AHK01}.

Several types of tiny
barrier inhomogeneities can be responsible for the large
current inhomogeneities which are observed in both C-AFM and numerical 
simulations \cite{TsP98}. First, due to the amorphous nature of
most insulator barriers, the metal-oxide interfaces can not be
perfectly smooth and fluctuations of the {\it barrier thickness} are
unavoidable. What is of interest here is the 'roughness'
of the barrier thickness which is usually much smaller than the
topography roughness \cite{DBB98,DTD00}. Thickness roughness of 
less than 0.1 nm can generate the fluctuations visible in 
\fig{fig1}b. There must also exist
inhomogeneities of the {\it barrier height}. Indeed changing a single
atom at a metal-oxide interface can induce local barrier changes
larger than 1 eV \cite{Sto02}. 
Statistics on the barrier parameters obtained from
many local C-AFM $I-V$ curves seem to indicate that barrier height
inhomogeneities play a more important role than thickness
inhomogeneities \cite{AKK00}. At last, metal oxides are known for
containing {\it electron traps} which are clearly evidenced in noise
studies \cite{BRG97} (see also \Section{s5}) and impurity states
\cite{WKU01,TsP01}. Further studies (C-AFM, physico-chemistry of
oxides, electronic and structural simulation, ballistic electron 
emission spectroscopy \cite{RPB01}...) are needed to
clarify the origin of current inhomogeneities.

\begin{figure}
\includegraphics[scale=0.33,angle=-90]{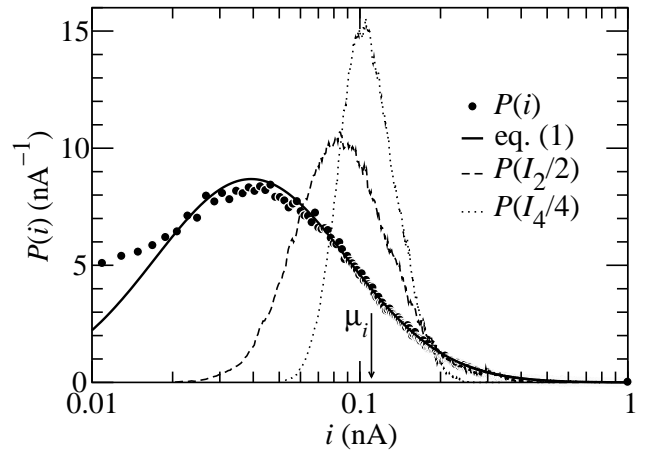}
\caption{Probability distribution $P(i)$ of the tunnel current
of \fig{fig1}b (open circles). The lognormal fit of the high current
tail of these data ($\mu=-2.5$, $\sigma=0.83$) is shown as solid
line (the small current tail is distorted by spurious noise). Also
indicated are the distributions $P(I_n/n)$ of currents flowing
through groups of $n$ shuffled pixels (\Section{s5}), which peak around the 
average current $\mu_i$ for large $n$.}
\label{fig2}
\end{figure}

Even without knowing the origin of the current inhomogeneities,
important consequences can be drawn from the knowledge of the
statistical distribution $P(i)$ of currents. The distribution
$P(i)$ presents in many cases \cite{DBB98,DHB00} a lognormal shape
(\fig{fig2}):
\begin{equation}
P(i) = \frac{1}{\sqrt{2\pi}\sigma i} \exp\left[ - \frac{\left( \ln
i -\mu \right)^2}{2\sigma^2} \right] = \LN(\mu,\sigma^2)(i),
\label{e3.1}
\end{equation}
where $\mu$ is a scale parameter and $\sigma$ is a shape
parameter, hereafter called the 'disorder strength'. Note that
standard $\chi^2$ fitting procedures are not adequate to fit
lognormal distributions because most of the current is usually
carried by the tail of the distribution while $\chi^2$ fitting
weights heavily the peak of the distribution, which does not carry
here a significant current and is strongly affected by spurious
noise. To take into account the large value tail more properly,
one can, \eg fit a parabola to the large $\ln i$ branch of $\ln
P(i)$ \vs $\ln i$ (a lognormal distribution is a parabola in
log-log scale).

The occurrence of lognormal distribution of tunnel currents is not
a surprise \cite{Bar97}. Suppose, indeed, that the distribution
$P(d)$ of barrier thickness $d$ is Gaussian with mean $\mu_d$ and
standard deviation $\sigma_d$. The current $i$ varies typically
exponentially with $d$ :
\begin{equation}
i = i_0 e^{-d/\lambda},
\end{equation}
where $i_0$ is a current scale and $\lambda$ is the attenuation
length of the electronic wave functions in the barrier. By
definition, the exponential of a Gaussian random variable is a
lognormal random variable. Thus, the current $i$ has a lognormal
distribution $\mathrm{LN}(\mu,\sigma^2)$ with $\mu = \ln i_0 -
\mu_d / \lambda$ and $\sigma = \sigma_d/\lambda$ \cite{Bar97}.
Importantly, the disorder strength $\sigma$ is $\sigma_d/\lambda$
and not $\sigma_d / \mu_d$ as one might expect naively. Thus, a
barrier which appears geometrically smooth ($\sigma_d/ \mu_d \ll
1$) might be 'rough' ($\sigma=\sigma_d/\lambda \gtrsim 1$) from
the point of view of current statistics, since typically $\lambda
\ll \mu_d$ ($\lambda \simeq 0.05-0.1$~nm, $\mu_d \simeq 1-2$~nm),
and generate large current inhomogeneities.

More generally, the tunnel current $i$ depends on the barrier
parameters $p_b$ (thickness, height, voltage, ...) typically as:
\begin{equation}
i = g_b \left( p_b \right) \exp \left[f_b \left( p_b \right) \right]
\end{equation}
where $g_b(p_b)$ and $f_b(p_b)$ vary less strongly than an
exponential. If $p_{b}$ presents small Gaussian fluctuations of
standard deviation $\sigma_{p_b}$ ($\sigma_{p_b} \ll \mu_{p_b})$
around its average value $\mu_{p_b}$, then one has $p_b =
\mu_{p_b} + \epsilon \sigma_{p_b}$ where
$\epsilon$ is a Gaussian random variable of order 1 (mean $0$,
standard deviation = 1). As $\sigma_{p_b} \ll \mu_{p_b}$, one can
write $f_b(p_b) = f_b(\mu_{p_b}) + \epsilon \sigma_{p_b}
f'_b(\mu_{p_b})$ and thus
\begin{equation}
i = g_b \left( p_b \right) \exp \left[ f_b \left( \mu_{p_b} \right) \right]
    \exp \left[ \epsilon \sigma_{p_b} f'_b \left( \mu_{p_b} 
    \right) \right].
\end{equation}
As $g_b(p_b)$ varies slowly, one can neglect its fluctuations.
Thus, $i$ appears as the product of a fixed term, $g_b \left( p_b \right) \exp
\left[ f_b \left( \mu_{p_b} \right) \right]$, by the exponential of a Gaussian
random variable, $\epsilon \sigma_{p_b} f'_b\left( \mu_{p_b} \right)$, 
which is, by definition, a lognormal random variable.

Thus lognormal distributions of tunnel currents emerge as the
consequence of \textit{small} Gaussian fluctuations of the
tunnelling parameters and are a good starting point to investigate the
tunnel current statistics.

\section{Simple consequences of the lognormal model}
\label{s4}

The lognormal distribution of currents (\eq{e3.1}) gives rise to peculiar
statistical properties. If the disorder strength is small ($\sigma \ll 1$),
the lognormal is close to a Gaussian and the usual
statistical behaviours, related to narrow distributions, appear. On the
contrary, if $\sigma$ is on the order of 1 or larger, the lognormal
distribution is broad and it presents a long tail, just as a L\'evy flight,
even if, unlike a L\'evy flight, it has finite average and
standard deviation.
The broadness of the lognormal also appears in the fact that
the typical current $\ityp$ (\ie most probable value),
\begin{equation}
\ityp = e^{\mu -\sigma^2},
\label{e4.1}
\end{equation}
can be much smaller than the average current $\mu_i$,
\begin{equation}
\mu_i = e^{\mu + \sigma^2/2},
\label{e4.2}
\end{equation}
indicating a large dispersion.

\begin{figure}
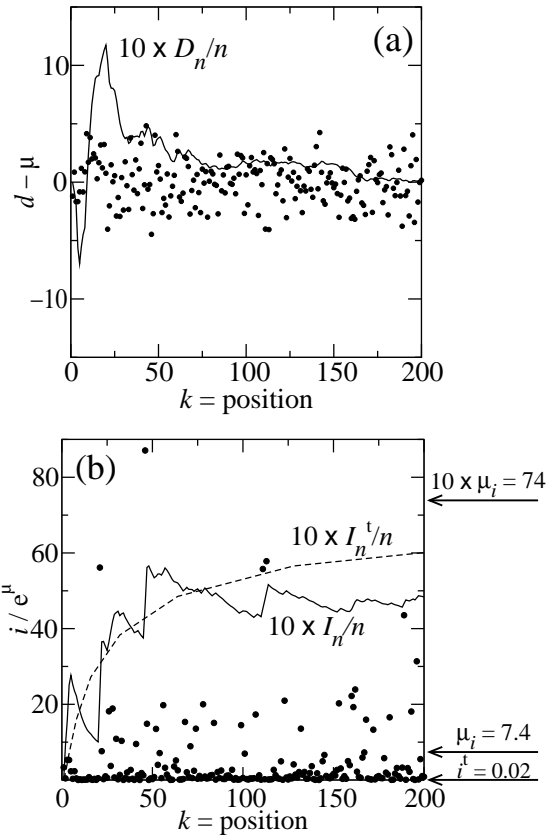

\includegraphics[scale=0.27,angle=-90]{fig3a.eps}
\includegraphics[scale=0.27,angle=-90]{fig3b.eps}
\caption{Differences between narrow and broad distributions.
Dots represent random values drawn from the distributions
described in the text; solid lines represent the partial sums of
these random values ($\times 10$), see \Section{s4}. (a) Gaussian
(narrow) distribution (b) lognormal (broad) distribution. The
broken line in (b) gives the typical scaling behaviour (see
\Section{s5}).}
\label{fig3}
\end{figure}

\Fig{fig3} illustrates the differences between narrow and broad
distributions. \Fig{fig3}a represents random values of a narrow
distribution, a Gaussian of arbitrary mean $\mu$ and standard
deviation $\sigma = 2$. The $X$-coordinate may represent the
position $k$ in a 1D tunnel barrier while the $Y$-coordinate may
represent the thickness or height of the barrier. All values are
of the same order of magnitude, $\mu$, to within about $\sigma$.
For the Gaussian, the typical value and the mean are equal. 
\Fig{fig3}b represents random values of a broad distribution, which is the
lognormal arising from the exponential of the Gaussian values of
\fig{fig3}a. \Fig{fig3}b may represent tunnel currents. The appearance
of the fluctuations is now completely different. The possible
values cover several orders of magnitude. Neither the typical
value nor the mean, which differ by a factor of 400, characterize
well the range of possible values. Thus, practically, the tunnel
current through a disordered barrier is not well characterized by
a single value like its mean but rather by the full distribution
(two parameters for a lognormal).

The large values observed in \fig{fig3}b correspond to the {\it
hotspots} in tunnelling. These are {\it not} pinholes: the large
currents arise from tiny fluctuations of barrier parameters 
(\fig{fig3}a) because the exponential dependence acts as a 'fluctuation
amplifier'. There is a {\it continuum} of large current values
corresponding to sites that have nothing qualitatively special,
but rather present small quantitative fluctuations of the barrier
parameters. With the lognormal model, one can estimate the current
inhomogeneity \cite{RDB02} by calculating the proportion $p_A$ of the
surface ($p_A = \int_{i_\alpha}^{\infty} P(i) \diff i$ where
$i_\alpha$ is a parameter) carrying the proportion $p_i$ of the
average current $\iave$ ($p_i = \int_{i_\alpha}^{\infty} i P(i)
\diff i / \iave$). For small $\sigma$ ($\sigma <0.25$), half of
the total current is carried by roughly half of the sites (no hot
spots). For large $\sigma$, the current
proportion carried by hot spots becomes more and more important.
For instance, for $\sigma =1$ (respectively 2), $p_i= 50\%$ of the
total current is carried on average by $p_A = 16\%$ (respectively
3\%) of the sites with highest current.

The domination of the current by the few largest transmission sites is
indirectly confirmed by the time fluctuations of the total current.
In certain conditions (small enough conditions, low temperatures), the current
exhibits strong telegraph noise indicative of electron trapping and detrapping
on a {\it single} trap in the barrier \cite{RSJ84,BRG97,NMW98}.
The large effect of a single
electron trap on the total current flowing through a junction is a strong clue
for the dominant role of a few hot spots.

At last, we comment \eq{e4.2} for the average current. One should
first be cautious that, contrary to what is frequently assumed
\cite{Cho63,Hur66}, the average $\mu_i$ is {\it not} in general
what is measured on a single tunnel junctions (see
Section~\ref{s5}). For a perfect barrier ($\sigma = 0$), we
recover the current $e^\mu$ without inhomogeneities, as expected.
The inhomogeneities generate a correcting 'disorder term',
$e^{\sigma^2/2}$. This term is always larger than 1: the barrier
inhomogeneities always {\it increase} the average current, due to
the non-linear dependence of the current on the fluctuating
parameters. Thus, the average current $\mu_i$ flowing through an
inhomogeneous barrier of given average thickness corresponds to
the current $i(d_\mathrm{eff})$ flowing through a {\it thinner}
effective homogeneous barrier ($d_\mathrm{eff}< \mu_d$), as
already noticed in \cite{Hur66}. Yet, in the model based on
thickness fluctuations, as $\mu = \ln i_0 - \mu_d /\lambda$, the
average current $\mu_i$ still depends exponentially on the average
thickness $\mu_d$, just as the current for a homogeneous barrier.
This is contrast with the modification of the $I-V$
characteristics shape by the presence of disorder \cite{Hur66}.

\section{Scale effects}
\label{s5}

The broad distribution of currents affects the size dependence of
tunnel junctions giving rise to anomalous scaling laws. To
understand this intuitively, we have plotted in figs. \ref{fig3}a and 
\ref{fig3}b
(solid lines) the quantities
\begin{equation}
D_n /n = \sum_{k = 1}^{n}{d_k} / n
\end{equation}
and
\begin{equation}
I_n/n = \sum_{k = 1}^{n}{i_k} / n
\end{equation}
which represent,
physically, the measured quantities at scale $n$. For instance
$I_n/n$ is proportional to the current per unit area flowing
through a junction of size $n$. For the Gaussian variable $d$,
$D_n /n$ is statistically distributed around the mean $\mu$ and
statistically converges to $\mu$ as $1/\sqrt{n}$ when $n$
increases (central limit theorem). At any scale, we have $D_n
\propto n$. For the lognormal variable $i$, the sum $I_n$ behaves
completely differently. At small scales, $I_n$ takes small values
very different from the mean and, as $n$ increases, there is a
slow upward trend of $I_n$ towards the mean $\mu_i$: this is the
anomalous scaling we investigate here. This upward trend is
created by the higher probability of larger samples (larger $n$'s)
to have a very large current peak $i_k$ that will significantly
draw the sum $I_n$ towards larger values. This effect does not
occur with narrow distributions like Gaussians for which the
largest terms in a statistical sample are not large enough to
modify the sums significantly.

These scale effects can be studied theoretically \cite{Bar97,RDB02}.
The problem reduces to finding the distribution of the sum  $I_n$ of
$n$ independent currents $i_k$ with the same lognormal distribution
$P(i) = \mathrm{LN}(\mu, \sigma^2)(i)$.
 For moderately broad lognormal distributions, we find
that $I_n$ is also approximately lognormally distributed, as
$\mathrm{LN}(\mu_n, \sigma_n^2)$, and has a typical value
\begin{equation}
\Int \simeq n \mu_i \left(1+C^2/n\right)^{-3/2}
\label{e5.1}
\end{equation}
where $\mu_i$ is the average current (\eq{e4.2}) and $C^2=e^{\sigma^2}-1$
is the coefficient of variation. The typical current is thus the product
of the usual term $n \mu_i$ by a correction term $\left(1+C^2/n\right)^{-3/2}$.
For small junctions ($n\ll C^2$), the correction term is important. Above
a characteristic size $n_\mathrm{c}=C^2$ related to the disorder strength,
the correction term tends slowly to 1 and the usual behaviour
$n \mu_i$ related to the law of large numbers is recovered.

\begin{figure}
\vspace*{5mm}
\includegraphics[scale=0.32,angle=-90]{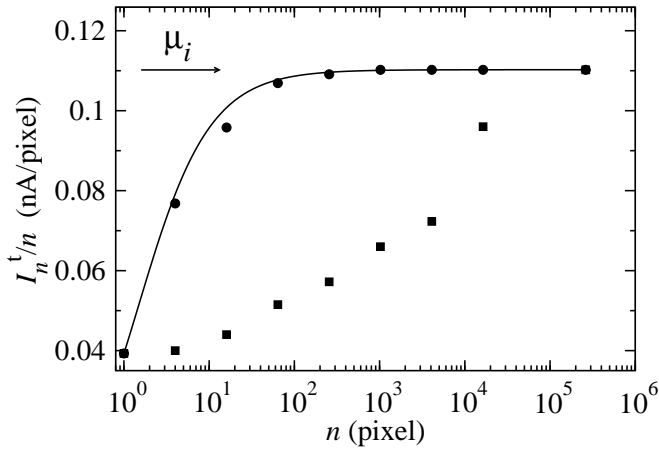}
\caption{Size dependence of the typical tunnel current density
$\Int/n$ of the \fig{fig1} junction. Squares (respectively circles)
corresponds to unshuffled pixels (respectively shuffled). The
solid line is the theoretical size dependence. The pixel area is
0.15 nm$^2$. However, one can not rigorously convert junction
sizes from pixels to nm$^2$ because the effective contact area of
the C-AFM tip is unknown.}
\label{fig5}
\end{figure}

The scaling relation \eq{e5.1} can be tested experimentally with
the current image of \fig{fig1}. For each $n$, we first sum the
currents of groups of $n$ pixels to obtain a statistical ensemble
of values $I_n$ and then construct the histograms presented in
\fig{fig2}. The histograms' peaks give $\Int$ (one must {\it not}
compute the mean but the typical value of $I_n$'s: the mean
presents no special scaling behaviour). Before grouping them, the
pixels have been spatially randomized to satisfy the condition of
statistical independence of the $i_k$'s. The result of this
procedure for $\Int/n$ is plotted in \fig{fig5}. As predicted, $\Int
/n$ deviates strongly at small scales from the constant $\mu_i$
that one would expect naively and the deviation is well described
by \eq{e5.1}. The maximum deviation is a factor $2.8 \simeq
e^{3\sigma^2 /2}$ for this good quality junction ($\sigma =
0.83$). For poorer quality junctions, deviations larger than
$10^2$ have been observed \cite{DHB00}.

If one takes into account the spatial correlations existing in the
barrier by not randomizing the pixels (black squares in \fig{fig5}),
there is still an anomalous scaling of $\Int /n$ and the
convergence towards $\mu_i$ is much slower than without
correlations. Thus the correlations play a crucial role in the
statistical properties of tunnel junctions as they combine with
the broadness of the current distribution to yield typical
currents differing from $n\mu_i$ even for relatively large sizes.

The size dependence predicted by \eq{e5.1} can also be tested by
measuring the currents of many patterned junctions of different
sizes to obtain the typical current at these sizes. This has been
done for semi-conducting AlAs barriers embedded in GaAs
\cite{Kel99}. Again, the typical current per unit size $\Int/n$ is
found to increase with the junction size, in qualitative agreement
with \eq{e5.1}.

In the tunnel junction community, size dependences are frequently
analyzed in terms of the product $R\times A$ of the resistance $R$
by the junction area $A$ and one usually checks that $R\times A$
does not depend on $A$, which is the normal size dependence.
However, anomalous size dependences have been reported recently
\cite{LWP02}. For 1~nm thick AlO$_x$ barriers, a significant {\it
increase} of $R\times A$ is found, from 60~$\Omega\mu$m$^2$ for $A
= 4~\mu$m$^2$ to 330~$\Omega\mu$m$^2$ for $A = 80~\mu$m$^2$. As
$R \propto 1/I_n$ and $A \propto n$, one has $R\times A \propto n/I_n$.
\Eq{e5.1} predicts an increase of $\Int/n$ with $n$ and thus, one expects
intuitively a decrease of $R\times A$ with $A$. Therefore, the
results of \cite{LWP02} apparently jeopardize \eq{e5.1}. However,
the correct theory \cite{RDB02} predicts that both the typical 
$I_n/n$ {\it and} the typical $n /I_n \propto R \times A$ increase 
with the junction size: 
\begin{equation}
(R\times A)^\mathrm{t} \propto (1+C^2/n)^{-1/2}, \label{e5.2}
\end{equation}
which shows how counter-intuitive broad distributions can be.

\section{Conclusions and overview}
\label{s6}

Several types of experiments (conductive AFM, noise studies,
scaling studies) provide a body of evidence that the distribution
of tunnel currents flowing through MIM junctions is broad, even in
good junctions. This is a natural consequence of the exponential
dependence of quantum tunnelling with the parameters. With such
broad distributions, the typical value is much smaller than the
average value so that tunnel currents should not be characterized
by a single number but instead by the full statistical
distribution. The lognormal distribution is found to fit the
experimental data in many cases. The shape parameter $\sigma$ of
lognormal distributions is a convenient figure of merit to compare
the quality of different junctions.

The broad character of the current distribution has several
implications, some obvious, some less obvious. First, the current
flows heterogeneously through the junction, in a way that the
lognormal model can quantify. Second, large {\it spatial}
variations of the current imply large {\it time} variations, \ie
large noise. Third, the average current varies strongly, typically
as $e^{\sigma^2}$ with the strength $\sigma$ of the disorder.
Fourth, the size dependence of the {\it typical} properties of
tunnel junctions (resistance or conductance) is affected
by the disorder, especially at small scales. One recovers the
usual size dependences at large scales but the transition from
small scale behaviour to large scale behaviour is slow. This
transition is further slowed down by spatial correlations.
Correspondingly, the large current inhomogeneities that exist at
small scale average out slowly when increasing the size of the
system. Thus, even relatively large junctions exhibit large
dispersions of conductances which are the relics of the poorly
averaged small scale inhomogeneities.

For applications, it is worth mentioning that the effects of disorder
increase rapidly when decreasing the junction size below a
characteristic size related to the disorder strength and to the spatial
correlations. To achieve better reproducibility, apart from the obvious
reduction of the barrier disorder, one could also aim at reducing the
spatial correlations.

{\small We thank K. Ounadjela who gave the impetus for our studies
of disorder in tunnel junctions and M. Chshiev, Y. Henry, A.
Iovan, M. Kelly and D. Stoeffler for discussions. This work was
partially supported by project 'NanoMEM' IST-1999-13471.}

\bibliographystyle{apsrev}
\bibliography{bibFB}

\newpage

\end{document}